\documentclass[onecolumn,fleqn,usenatbib]{mnras}

\usepackage[T1]{fontenc}
\usepackage{ae,aecompl}
\usepackage{hyperref}

\usepackage{graphicx}    
\usepackage{amsmath}    
\usepackage{amssymb}    

\usepackage{orcidlink}
\usepackage{subfig}

\newcommand{\source}{1LHAASO J\ensuremath{1945+2424}~}

\title[A GeV counterpart to 1LHAASO J1945+2424]{Discovery of an extended GeV counterpart to the TeV source 1LHAASO J1945+2424 in \textit{Fermi}-LAT data}

\author[Araya \& \'Alvarez-Quesada]{
Miguel Araya,\,\orcidlink{0000-0002-0595-9267}$^{1}$\thanks{E-mail: miguel.araya@ucr.ac.cr},
J.A.~\'Alvarez-Quesada,$^{1}$
\\
$^{1}$Escuela de F\'isica, Universidad de Costa Rica, Montes de Oca, San Jos\'e, Costa Rica, 11501-2060}

\date{Accepted ; Received ; in original form }

\pubyear{2023}

\begin{document}
\label{firstpage}
\pagerange{\pageref{firstpage}--\pageref{lastpage}}
\maketitle

\begin{abstract}
With almost 15 years of data taken by the Large Area Telescope (LAT) onboard the \emph{Fermi} satellite we discovered an extended source of GeV emission in the region of the very-high-energy (TeV) source \source. This TeV source is more extended than the LAT source. The spectrum of the GeV emission is hard (with a photon spectral index $\sim 1.5$) and connects smoothly with that of the TeV source, indicating a likely common origin. In order to explain the origin of the $\gamma-$rays we explore scenarios which are typically used for supernova remnants (SNRs) and pulsar wind nebulae (PWN). For an SNR with a single particle population, a leptonic particle distribution in the form of a broken power-law with a break energy of $\sim 3.7$~TeV explains the spectra well, while in the hadronic scenario a simple power-law with a hard spectral index of $\sim1.64$ is necessary. In the PWN scenario, reasonable parameters are obtained for a source age of 10~kyr and current pulsar spin-down luminosity of $\sim 10^{34}$~erg~s$^{-1}$.
\end{abstract}


\section{Introduction}\label{sec:intro}
The Large High Altitude Air Shower Observatory (LHAASO) is a ground-based extensive air shower detector consisting of the 78,000~m$^2$ Water Cherenkov Detector Array (WCDA) for TeV astronomy and the 1.3~km$^2$ array (KM2A) for $\gamma-$ray detection above 10 TeV \citep{2022ChPhC..46c0001M}. Located in Daocheng, Sichuan province of China, these arrays detect events in the declination range from $-20\degr$ to $80\degr$, thus continuously scanning a large portion of the northern sky.

The first LHAASO catalog has been recently published \citep{2023arXiv230517030C}. It constains 32 new sources of TeV emission and details on many other previously known sources, including the extended source \source found in the plane of the Galaxy at a Galactic longitude of $\sim 60\degr$. WCDA detected this source in the $1-25$~TeV energy range which was modeled with a 2D gaussian morphology having a 39\%-containment radius ($r39$) of $1.29\pm 0.11\degr$. The coordinates (J2000) of the gaussian centre are RA$=296.36\degr$, Dec$=24.40\degr$ (positional uncertainty $0.37\degr$). Fitting a power-law model to the differential spectrum of \source, $\frac{dN}{dE}=N_0(\frac{E}{3\,\mbox{\tiny TeV}})^{-\Gamma}$, yielded a photon index $\Gamma=2.56\pm0.08$ and $N_0 = (4.27\pm 0.51)\times 10^{-13}$~cm$^{-2}$~s$^{-1}$~TeV$^{-1}$ \citep{2023arXiv230517030C}. Above an energy of 25~TeV, a source with a smaller angular size and a lower flux is detected by KM2A, centered about $1\degr$ to the southeast of the WCDA centroid coordinates. The KM2A source is modeled by a 2D gaussian with $r39 =0.36\pm0.06\degr$ and a much steeper power-law spectrum with an index of $3.93\pm0.30$. \cite{2023arXiv230517030C} studied the effect of the Galactic diffuse emission model on the source parameters (the extensions, locations and spectral parameters), particularly for the extended sources, and found that the properties measured by WCDA and KM2A for \source were not affected by excluding the Galactic diffuse emission. However, due to their morphological and spectral differences, it is not clear whether the components seen by WCDA and KM2A belong to the same object.

Although there are no previous reports of an extended $\gamma-$ray source consistent with the location of \source, initial observations of the region by the High Altitude Water Cherenkov \citep[HAWC,][]{2017ApJ...843...40A} observatory, which also scans a large portion of the northern TeV sky continuously \citep{2023NIMPA105268253A}, have found several TeV sources. HAWC's angular resolution ranges between $0.1\degr$ and $1.0\degr$ (68\%-containment radius for photons) depending on the energy and zenith angle of the signal \citep{2020ApJ...905...76A}. The source 2HWC~J$1938+238$ was found in a point source search. This source is located $\sim1.6\degr$ to the west of the WCDA centroid position of \source. Another possibly extended source, labeled 2HWC~J$1949+244$, and located $\sim 2\degr$ to the southeast of the WCDA centroid of \source was found in their extended source search. More recent observations by HAWC \citep{2020ApJ...905...76A} have revealed only two dim point sources (and no extended source) in the region, 3HWC~J$1950+242$ to the east of the WCDA emission and located within the 39\%-containment radii of both the KM2A and WCDA sources. Its positional statistical uncertainty is $0.11\degr$. The other point source reported was 3HWC~J$1940+237$, seen close to the location of 2HWC~J$1938+238$, and with a relatively large positional statistical uncertainty of $0.36\degr$. None of these TeV sources have known counterparts at lower energies.

Regarding observations at GeV energies, four point sources are found in the \textit{Fermi} Large Area Telescope Fourth Source Catalog within an angular distance of $r39$ from the WCDA source \citep{2020ApJS..247...33A}. These are 4FGL~J$1948.1+2511$ and 4FGL~J$1948.8+2420$, having no known associations, 4FGL~J$1950.6+2416$, associated to the millisecond pulsar PSR~J$1950+2414$, and 4FGL~J$1946.1+2436c$, possibly associated to a radio source of unknown type, NVGRC~J$194625.5+243638$ \citep{2020ApJS..247...33A}. According to the ATNF Pulsar Catalogue \citep{2005AJ....129.1993M} the pulsar PSR~J$1950+2414$ is located at a distance of 7.27~kpc and has an age of $3.63\times 10^9$~yr and a spin-down luminosity of $9.3\times 10^{33}$~erg~s$^{-1}$. All these 4FGL sources show spectra that decrease steeply at GeV energies, except for the dim source 4FGL~J$1948.8+2420$, with a spectrum described by a simple power law with a spectral index of $\sim2.2$.

Supernova remnants (SNRs), the remains of exploding stars, and pulsar wind nebulae (PWN), are known $\gamma-$ray sources resulting from the relativistic particles that these systems produce \citep[see, e.g.,][]{2016ApJS..224....8A,2017ApJ...843..139A}. These classes of objects constitute the majority of the identified Galactic TeV sources \citep{2018A&A...612A...1H}. $\gamma-$ray sources associated with these objects are often seen as extended, reaching up to several degrees in the sky \citep[e.g.,][]{2017ApJ...843..139A,2020A&A...643A..28D,2020MNRAS.492.5980A,2020ApJS..247...33A}. Two known SNRs are found to the west and southwest of the WCDA source, G$59.8+1.2$ and G$59.5+0.1$, according to the University of Manitoba's high-energy catalogue of SNRs \citep{2012AdSpR..49.1313F}\footnote{See \url{http://snrcat.physics.umanitoba.ca}}. These SNRs are located about $1.5\degr$ and $1.2\degr$ from the centre of \source and have small angular sizes compared to the extension of the TeV emission ($<20'$).

No PWN are known in the region. A search in the ATNF Pulsar Catalogue \citep{2005AJ....129.1993M} for pulsars located within $1.5\degr$ from the WCDA emission centroid results in 10 sources. The ones having the highest spin-down luminosities are PSR~J$1940+2337$ ($\dot{E}=1.9\times 10^{34}$~erg~s$^{-1}$, seen close to 3HWC~J$1940+237$, this pulsar is also the youngest with an age of $1.13\times10^5$~yr), the previously mentioned PSR~J$1950+2414$ ($\dot{E}=9.3\times 10^{33}$~erg~s$^{-1}$), found close to the location of 3HWC~J$1950+242$, and PSR~J$1948+2333$ ($\dot{E}=3.6\times 10^{33}$~erg~s$^{-1}$) seen towards the south of the KM2A source. The catalogued distances to these two pulsars are relatively large, 8.5, 7.27 and 8.0~kpc, respectively. From the pulsars in the vicinity, the closest is PSR~J$1946+24$, at a catalogued distance of $\sim 4.3$~kpc, although its age and spin-down luminosity have not been measured. However, with a period of 4.729~s \citep{2009ApJ...703.2259D} PSR~J$1946+24$ is likely to have a low spin-down luminosity.

Motivated by the unidentified nature of \source we searched for a GeV signal in the region using data from the \textit{Fermi}-LAT to try to constrain its properties. In section \ref{LATdata} we present the results of the data analysis which shows the existence of an extended source of GeV emission, and in section \ref{discussion} we model the spectrum of \source from GeV to TeV energies using simple models representing SNR and PWN scenarios.

\section{\textit{Fermi}-LAT Data}\label{LATdata}
The \emph{Fermi}-LAT is a converter/tracker telescope detecting $\gamma-$rays in the energy range between 20~MeV and $\ga$1~TeV \citep{2009ApJ...697.1071A}. We gathered Pass 8 data from 27 October 2008 to 30 June 2023, or a Mission Elapsed Time (MET) period $246823875-709833870$, in the energy range 200~MeV--500~GeV. The point spread function (PSF) of the telescope (68\% containment angle) for front+back-converted events is $\sim0.8^\circ$ at 1~GeV and drops below $\sim0.15^\circ$ above 10~GeV, compared to $\sim5^\circ$ at 100~MeV. Data taken in the first two months of the mission, prior to MET 246823875, were not used due to high levels of background contamination above $\sim 30$~GeV\footnote{See \url{https://fermi.gsfc.nasa.gov/ssc/data/analysis/LAT\_caveats.html}}.

The data were analyzed with the {\tt fermitools} version~2.2.0 by means of the {\tt fermipy} package version~1.1.6 following the recommended cuts\footnote{See \url{https://fermi.gsfc.nasa.gov/ssc/data/analysis/documentation/Cicerone/Cicerone\_Data\_Exploration/Data\_preparation.html}}. We selected front and back-converted {\tt SOURCE} class events having good quality with the options {\tt evclass=128}, {\tt evtype=3}, {\tt DATA\_QUAL$>0$}, and with zenith angles lower than $90\degr$ to avoid contamination from $\gamma-$rays from Earth's limb. We used the response functions associated to the dataset, {\tt P8R3\_SOURCE\_V3}, and we binned the data with a spatial scale of $0.05\degr$ per pixel and ten bins in energy for exposure calculations.

The analysis of LAT data to obtain the spectral and morphological parameters of a source relies in the maximum likelihood technique \citep{1996ApJ...461..396M} by which the parameters are adjusted to match the number of events in each spatial and energy bin. Given the relatively large PSF we analyzed a region of interest (RoI) in the sky around the target source. The centre of the RoI was chosen at the coordinates (J2000) RA $= 296\degr$, Dec $= 23.9\degr$, and we used the events reconstructed within $15\degr$ of this location for the analysis. The model included the sources found within $20\degr$ of the RoI centre and are given by the data release 3 of the fourth catalog of LAT sources (4FGL-DR3), based on 12 years of survey data in the 50~MeV--1~TeV energy range \citep{2020ApJS..247...33A,2022ApJS..260...53A}, as well as the Galactic diffuse emission model (using the file {\tt gll\_iem\_v07.fits}) and the isotropic and residual cosmic-ray background (modeled by {\tt iso\_P8R3\_SOURCE\_V3\_v1.txt})\footnote{See \url{https://fermi.gsfc.nasa.gov/ssc/data/access/lat/BackgroundModels.html}}. The energy dispersion correction was applied to all sources except for the isotropic component, as recommended by the LAT team\footnote{See \url{https://fermi.gsfc.nasa.gov/ssc/data/analysis/documentation/Pass8\_edisp\_usage.html}}.

In the maximum likelihood framework the detection significance of a source is obtained from its test statistic (TS), defined as $-2\log(\mathcal{L}_0/\mathcal{L})$, where $\mathcal{L}$ is the maximum likelihood for a model including a new source and $\mathcal{L}_0$ is the maximum likelihood for a model without the source (the null hypothesis). In all the analyses we left the spectral normalizations of the sources located within $10\degr$ of the RoI centre free to vary as well as the spectral indices and normalizations of the sources located within $5\degr$ of the centre. We fixed the parameters of the rest of the sources to the values reported in the catalog.

\subsection{Morphology}
Given that \source is a very-high-energy source whose spectral energy distribution (SED, $E^2\frac{dN}{dE}$, with $\frac{dN}{dE}$ the differential energy spectrum) is known to decrease as a function of energy above an energy of 1~TeV, it is reasonable to search for a LAT counterpart at the highest energies. This also benefits from the improved resolution of the LAT, which is appropriate for morphological studies, and also lowers the effect of the Galactic diffuse emission and the background sources on the analysis. We found that above an energy of 20~GeV none of the 4FGL sources found within $2.2\degr$ of the centre of the RoI, 4FGL~J$1946.1+2436c$, 4FGL~J$1950.6+2416$, 4FGL~J$1948.1+2511$ and 4FGL~J$1951.0+2523$, were significantly detected (TS$\sim 0$), except for the source 4FGL~J$1948.8+2420$, which is detected with a significance of $\sim 4\sigma$. The undetected sources have spectra described by log-parabolas in the 4FGL catalog, with spectral indices greater than $\sim 2.5$, which explains why we did not detect them at high energies. The spectral shapes of these sources is confirmed in section \ref{spectrum} below. As mentioned earlier the source 4FGL J$1950.6+2416$ is associated to the millisecond pulsar PSR J$1950+2414$ and in that case it is expected for the source to not show up in the maps at the highest energies. The other sources, however, have no known associations.

\begin{figure}
    \centering
    \includegraphics[width=0.47\textwidth]{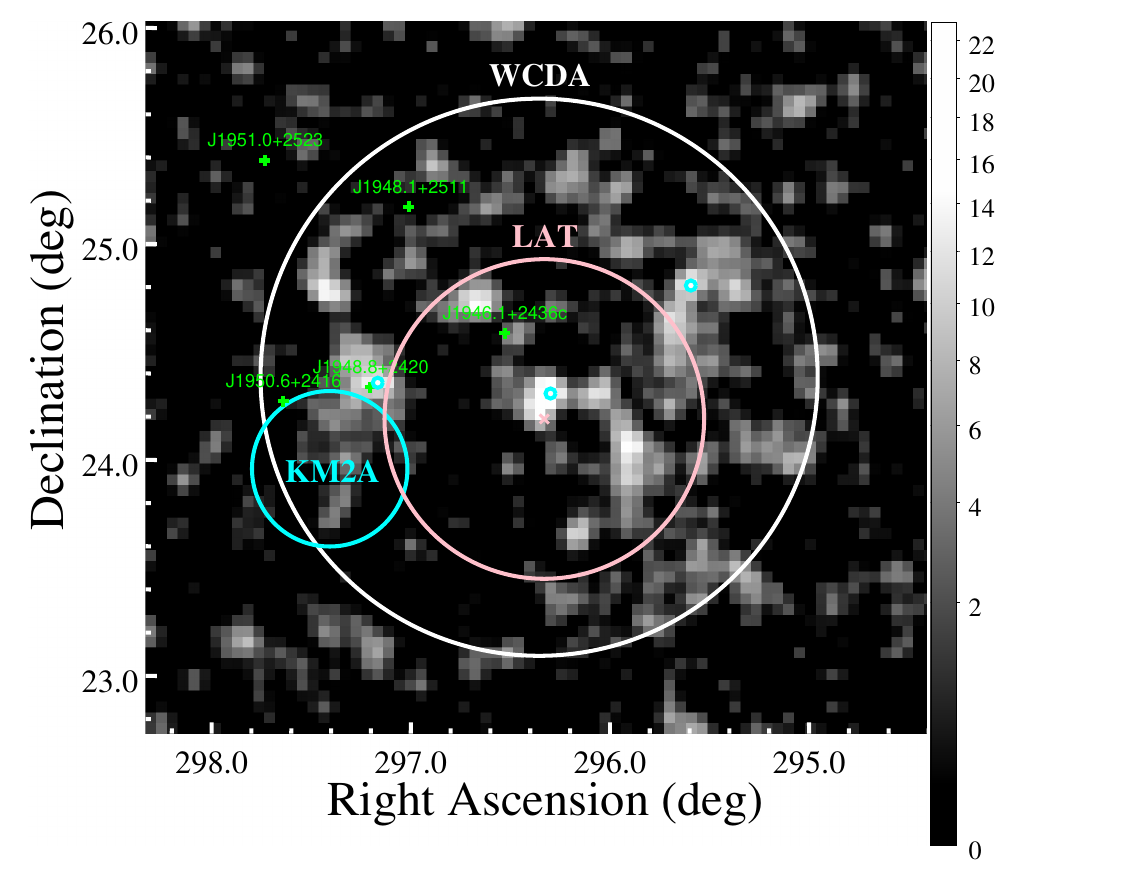}
    \includegraphics[width=0.47\textwidth]{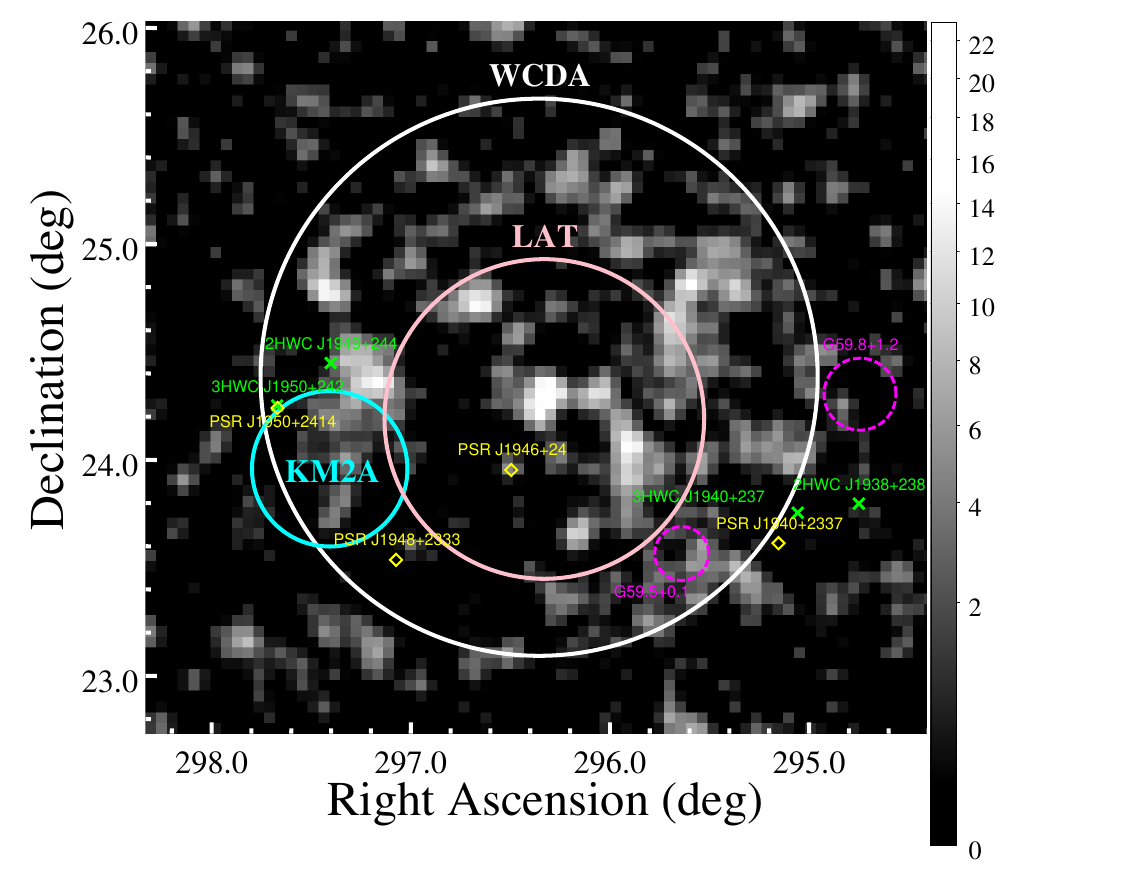}
    \caption{LAT TS map above an energy of 20~GeV showing the newly found emission region. The gray scale bars show the TS values. \textit{Left:} Large circles represent the 39\%-containment radii of the 2D gaussian models for the LAT, WCDA and KM2A sources. The `x' marks the centroid of the LAT emission and the locations of the 4FGL sources are labeled, of which only 4FGL~J$1948.8+2420$ is detected as an excess. The small circles represent the locations of TS maxima found with events having energies above 20~GeV in a point source search (the point sources in Table \ref{table1}, see text). \textit{Right:} The locations of the pulsars (boxes), SNRs (dashed circles) and HAWC sources (`x') mentioned in section \ref{sec:intro} are shown together with the LHAASO sources and the GeV source found in this work}.
    \label{fig:tsmap}
\end{figure}

After removing the sources 4FGL~J$1946.1+2436c$, 4FGL~J$1950.6+2416$, 4FGL~J$1948.1+2511$, 4FGL~J$1951.0+2523$ and 4FGL~J$1948.8+2420$ from the model we obtained a TS map by fitting the normalization of a test point source located at every pixel in the map assuming a simple power-law with an index of 2 to calculate its TS value, which is then represented in the map. The TS map allows to probe for the existence of new sources not present in the model. Figure \ref{fig:tsmap} shows the resulting map as well as the locations and extensions of the LHAASO sources and the locations of the 4FGL sources. The figure also shows in a separate diagram the locations of the pulsars, the SNRs and the catalogued HAWC sources discussed in section \ref{sec:intro}. A $\sim4\sigma$ excess is found at the location of 4FGL~J$1948.8+2420$ as well as at other nearby locations within the extent of \source.

We searched for new sources in the RoI by running the tool {\tt find\_sources}. This algorithm calculates a TS map of the RoI by fitting the normalization of a test point source located at each pixel assuming a fixed spectral shape, which we chose to be a simple power-law with index 2. For each peak in the map having a value above a threshold defined by the user (which we set to 16) a point source is added to the model iteratively and its location is optimized by fitting a two-dimensional parabola to the log-likelihood surface around the peak maximum. After each source is added its spectral parameters are fitted (again assuming a simple power-law spectral shape) and the sources with their final spectral parameter values are incorporated in the model. We found ten new point sources, three of which are located within the 39\%-containment region of the WCDA source. We labeled these sources as PS~J$1945.2+2419$, PS~J$1942.3+2449$ and PS~J$1948.7+2422$, the latter is consistent with the location of 4FGL~J$1948.8+2420$, and show them in figure \ref{fig:tsmap}. The spectral indices $\Gamma$ and normalizations $N_0$ of these sources are defined in the function $\frac{dN}{dE} = N_0\left(\frac{E}{1000\,\mbox{\tiny MeV}}\right)^{-\Gamma}$ and are summarised in Table \ref{table1}.

The fact that the newly found point sources in the region of \source all show similar hard GeV spectral indices, and flux normalizations that are consistent within the uncertainties between the sources indicates that they might instead be part of the same extended source. We replaced the three point sources with an extended source modeled by a 2D gaussian \citep[for its definition, see][]{2012ApJ...756....5L}, as done in the LHAASO search for extended sources, and having a spectral shape of the form $\frac{dN}{dE} = N_0\left(\frac{E}{1000\,\mbox{\tiny MeV}}\right)^{-\Gamma}$. We ran a scan in parameter space to maximize the likelihood and thus obtain the best-fit extension and centroid location (and their statistical uncertainties), which are $r39 = 0.74^{+0.08}_{-0.06}\degr$, RA $= 296.3\pm0.2\degr$, Dec $= 24.2\pm0.3\degr$ (J2000). Interestingly, the coordinates of the GeV centroid are fully consistent with those of the WCDA source although the extension of the LAT source is smaller ($r39=1.29\pm 0.11\degr$ for the TeV source). The best-fit spectral parameters of the extended source are also indicated in Table \ref{table1} and its extension ($r39$) is indicated in figure \ref{fig:tsmap}.

\begin{table*}
\centering
\caption{Spectral parameters of the sources found in the morphological analysis with energies above 20~GeV ($1\sigma$ statistical uncertainties quoted)}
\begin{tabular}{|c|c|c|c|c|}
\hline
\hline
Spatial model & & $\Gamma$ & $N_0$ ($10^{-13}$~MeV$^{-1}$~cm$^{-2}$~s$^{-1}$) & TS\\
\hline
 & PS~J$1945.2+2419$ & $1.33\pm0.06$ & $0.24\pm 0.08$ & 27.9\\
PS & PS~J$1942.3+2449$ & $1.36\pm0.06$ & $0.26\pm 0.10$ & 19.5\\
 & PS~J$1948.7+2422$ & $1.21\pm0.06$ & $0.11\pm 0.05$ & 19.9\\
\hline
2D Gaussian & & $1.86 \pm 0.10$ & $32\pm 14$ & 79.8\\
\hline
\end{tabular}\\
\label{table1}
\end{table*}

In the morphological analysis with energies above 20~GeV the difference in TS values for the best-fit gaussian and that of a single point source at an optimized location that maximizes the likelihood is TS$_{\mbox{\tiny ext}} = 46.2$. Studies of the distribution of this parameter have shown that a value TS$_{\mbox{\tiny ext}}=16$ is an appropriate lower threshold for choosing the extended template as a preferred model for a source \citep{2012ApJ...756....5L}. We also calculated the Akaike Information Criterion \citep{1974ITAC...19..716A} defined as AIC$= 2k - 2\ln(\mathcal{L})$, where $k$ is the number of free parameters in the model, thus taking into account the quality of the fit while also incorporating a penalty related to the amount of free parameters. We compared the AIC values for a model containing the three point sources within the extension of \source to that of the single extended source. The model with the lowest AIC, the preferred model, is the gaussian template having a lower AIC value by $\Delta$AIC$=26.6$.

\subsection{LAT spectrum}\label{spectrum}
With the spatial model found in the previous step we proceeded to obtain the spectrum of the source in the 0.2--500~GeV energy range. We added the gaussian template found in the previous section to the 4FGL model, replacing the point source 4FGL~J$1948.8+2420$. We used a simple power-law spectrum for the extended source and, after freeing the spectral parameters of the background sources as before, we performed a maximum likelihood fit. The sources 4FGL~J$1946.1+2436c$, 4FGL~J$1950.6+2416$, 4FGL~J$1948.1+2511$ and 4FGL~J$1951.0+2523$ are significantly detected and the TS value of the extended source is 83.3. We built SEDs for the 4FGL point sources and the extended source by dividing the energy range in ten intervals equally spaced logarithmically and fitting the normalization of the sources in each interval to obtain the flux points. We kept the spectral index of the sources fixed to the value obtained in the global fit (see below). When a source was not detected in a given interval (i.e., having TS~$<4$) we calculated a 95\%-confidence level (CL) upper limit on the flux. By inspecting the SEDs we concluded that the 4FGL sources are not significantly detected above several GeV, while the extended source is not detected below $\sim 2$~GeV. The resulting SED flux points of the extended source are shown in table \ref{SEDtable}.

We compared two fits using different spectral shapes for the extended source, a simple power-law of the form $\frac{dN}{dE}=N_0 \left( \frac{E}{E_0}\right)^{-\Gamma}$ and a log-parabola given by $\frac{dN}{dE}=N_0 \left( \frac{E}{E_0}\right)^{-(\alpha + \beta \,\mbox{log}(E/E_0))}$, where $E_0$ is a fixed scale factor. The parameters $N_0$, $\Gamma$, $\alpha$ and $\beta$ are left free to vary in each case. The differences in TS values obtained between the more complex model (log-parabola) and the simpler one (power-law) was $\Delta$TS$=14.9$ which, according to the likelihood-ratio test for nested models \citep{wilks1938}, indicates that the log-parabola is statistically preferred over the simple power-law at a $\sim3.9\sigma$ level. This shows some weak evidence of spectral curvature but given that its significance is below the $5\sigma$ level we adopt the simple power law for the spectrum.

The values of the normalization and spectral index for the extended source are $\Gamma=1.51\pm 0.02_{\mbox{\tiny stat}} \pm 0.35_{\mbox{\tiny sys}}$ and $N_0=(5.0\pm 0.2_{\mbox{\tiny stat}} \pm 0.7_{\mbox{\tiny sys}})\times 10^{-15}$~MeV$^{-1}$cm$^{-2}$s$^{-1}$, respectively, for $E_0=25000$~MeV. The systematic uncertainties shown are the errors in modeling the Galactic diffuse emission. To estimate them we modeled the diffuse emission with the eight alternative models explored by \cite{2016ApJS..224....8A}, scaled appropriately to account for the differences in energy dispersion between {\tt Pass 7} (for which they were developed originally) and {\tt Pass 8} reprocessed data\footnote{See \url{https://fermi.gsfc.nasa.gov/ssc/data/access/lat/Model\_details/Pass8\_rescaled\_model.html}}. We calculated the uncertainties in the spectral parameters as in \cite{2016ApJS..224....8A} from the set of alternative values. We detected the source significantly using all the alternative models (TS~$>50$).

\begin{table*}
\centering
\caption{SED flux points and TS values in energy intervals for the extended LAT source in the region of \source}
\begin{tabular}{|c|c|c|}
\hline
\hline
Energy range (GeV) & $E^2\frac{dN}{dE}$ ($10^{-7} $~MeV~cm$^{-2}$~s$^{-1}$)$^a$ & TS\\
\hline
0.2--0.44 & $1.8^b$ & -\\
0.44--0.96 & $5.4^b$ & -\\
0.96--2.1 & $9.0^b$ & -\\
2.1--4.6 & $19.6\pm7.7$ & 9\\
4.6--10 &$29.0\pm7.4$  & 12\\
10--22 & $43.5\pm9.4$ & 19\\
22--48 & $33.2\pm12$ & 12\\
48--105 & $77.8\pm15$ & 29\\
105--229 & $55.0\pm21$ & 13\\
229--500 & $72.9\pm26$ & 11\\
\hline
\end{tabular}\\
\textsuperscript{$a$}\footnotesize{$1\sigma$ statistical uncertainties are given for the fluxes.}\\
\textsuperscript{$b$}\footnotesize{95\%-CL upper limit on flux.}
\label{SEDtable}
\end{table*}

\section{Discussion}\label{discussion}
We found an extended GeV source at the location of the TeV LHAASO source \source. Although the WCDA component of the TeV emission is more extended than the LAT source the spectra measured by WCDA and \textit{Fermi}-LAT can be joined smoothly, as we will show here. The component detected by KM2A at energies above 25~TeV, on the other hand, is smaller than the WCDA and LAT sources, as can be seen in Fig. \ref{fig:tsmap}. In our models we assumed that all three components are related to the same source and did not attempt to explain the different morphologies, which is left for future work.

Examples of extended GeV and TeV sources of unknown nature showing hard GeV spectral indices, as is the case for \source, include 2HWC~J$2006+341$ \citep{2020ApJ...903L..14A}, FHES~J1741.6 -- 3917 \citep[labeled as G$350.6-4.7$ by][]{2018MNRAS.474..102A} and others \citep{2018ApJS..237...32A}. Other sources with similar GeV spectra have been recently associated to new supernova remnants, such as G$17.8+16.7$ \citep{2022MNRAS.510.2920A} and a source possibly associated to the Calvera pulsar \citep[labeled as G$118.4+37.0$ by][]{2023MNRAS.518.4132A}.

In this section we apply simple models to explain the spectrum of \source assuming that it is an SNR or a PWN and derive the values of the relevant physical parameters required.

\subsection{Supernova remnant hypothesis}
In figure \ref{fig:tsmap} we show the locations and extensions of the two known SNRs in the region of \source, G$59.5+0.1$ and G$59.8+1.2$. Given their small angular diameters compared to the extensions of the GeV and TeV sources we can rule out these SNRs as the cause of the $\gamma-$rays. We thus explored the properties of the known SNR population to try and constrain some of the parameters in the SNR hypothesis for the emission.

In the SNR population with known distances the largest objects show diameters of $\sim$100~pc \citep{2023ApJS..265...53R}, which is consistent with the typical size predicted by evolutionary models near the end of the Sedov-Taylor phase \citep{2017AJ....153..239L}. Assuming a physical radius of $r<50$~pc corresponds to the observed angular radius of $1.29\degr$ (the WCDA extension of \source) implies that the distance to the SNR is $d<2.2$~kpc. However evolved SNRs usually show GeV spectral indices greater than 2 \citep{2016ApJS..224....8A} meaning that \source has a LAT spectrum which is more typical of an SNR with an age $<10$~kyr. Adopting SNR radii of $12-20$~pc which are predicted for a 10$-$kyr old source with an initial explosion energy of $10^{51}$~erg and an ejecta mass of $M_\odot$ evolving in a homogeneous medium with densities in the range $0.1-1$~cm$^{-3}$ \citep{1999ApJS..120..299T}, we obtain distances of $0.5-0.8$~kpc. From the results of our LAT data analysis, these distances imply GeV luminosities ($1-100$~GeV) of $0.5-1.4\,\times10^{33}$~erg~s$^{-1}$ which are consistent with population trends \citep{2016ApJS..224....8A}. We adopt a distance of 0.65~kpc for the source in this section.

In an SNR the $\gamma-$rays can be produced by up-scattering low energy photons by relativistic electrons accelerated by SNR shocks (inverse Compton scattering, or IC). On the other hand, if relativistic nuclei are also present in the SNR, inelastic collisions between these cosmic rays and other nuclei in dense ambient regions can lead to the production of unstable neutral pions which decay into $\gamma$-rays. We explored these two scenarios and used the {\tt NAIMA} package \citep{Zabalza:2016mu} to fit the data points. For the LHAASO flux points we used the published flux values at the reference energies reported for both arrays ($3$~TeV for WCDA and $50$~TeV for KM2A).

The IC leptonic scenario for the $\gamma-$rays is a natural explanation for the GeV$-$TeV spectra given the hard GeV spectral slope followed by softening of the emission. We considered as target photons the cosmic microwave background (CMB), described by a blackbody photon field with an energy density of $0.26$~eV~cm$^{-3}$ and a temperature of $2.73$~K, as well as graybody \citep[i.e., a blackbody with a dilution factor, see][]{2014ApJ...783..100K} optical (NIR) and infrared (FIR) photon fields approximating the main componentes of the local interstellar fields. We found that the densities and temperatures for the NIR and FIR fields of $1.0$~eV~cm$^{-3}$ and $3000$~K, $0.5$~eV~cm$^{-3}$ and $30$~K, respectively, describe reasonably well the local interstellar field SED \citep[as seen in][]{2017MNRAS.470.2539P}, although our results depend weakly on the choice of these parameters. As an example we also show the fit values obtained after using the interstellar field parameters found by \cite{2018A&A...617A..78T} (FIR: $0.2$~eV~cm$^{-3}$ and $30$~K, NIR: $0.3$~eV~cm$^{-3}$ and $3000$~K). {\tt NAIMA} implements the IC calculation from \cite{2014ApJ...783..100K}. For synchrotron emission, the parametrization of the emissivity function presented by \cite{2010PhRvD..82d3002A} is adopted. Finally, the energy spectra and production rates of $\gamma-$rays from hadronic interactions is taken from \cite{2014PhRvD..90l3014K}.

\begin{figure}
    \centering
    \includegraphics[width=0.7\textwidth]{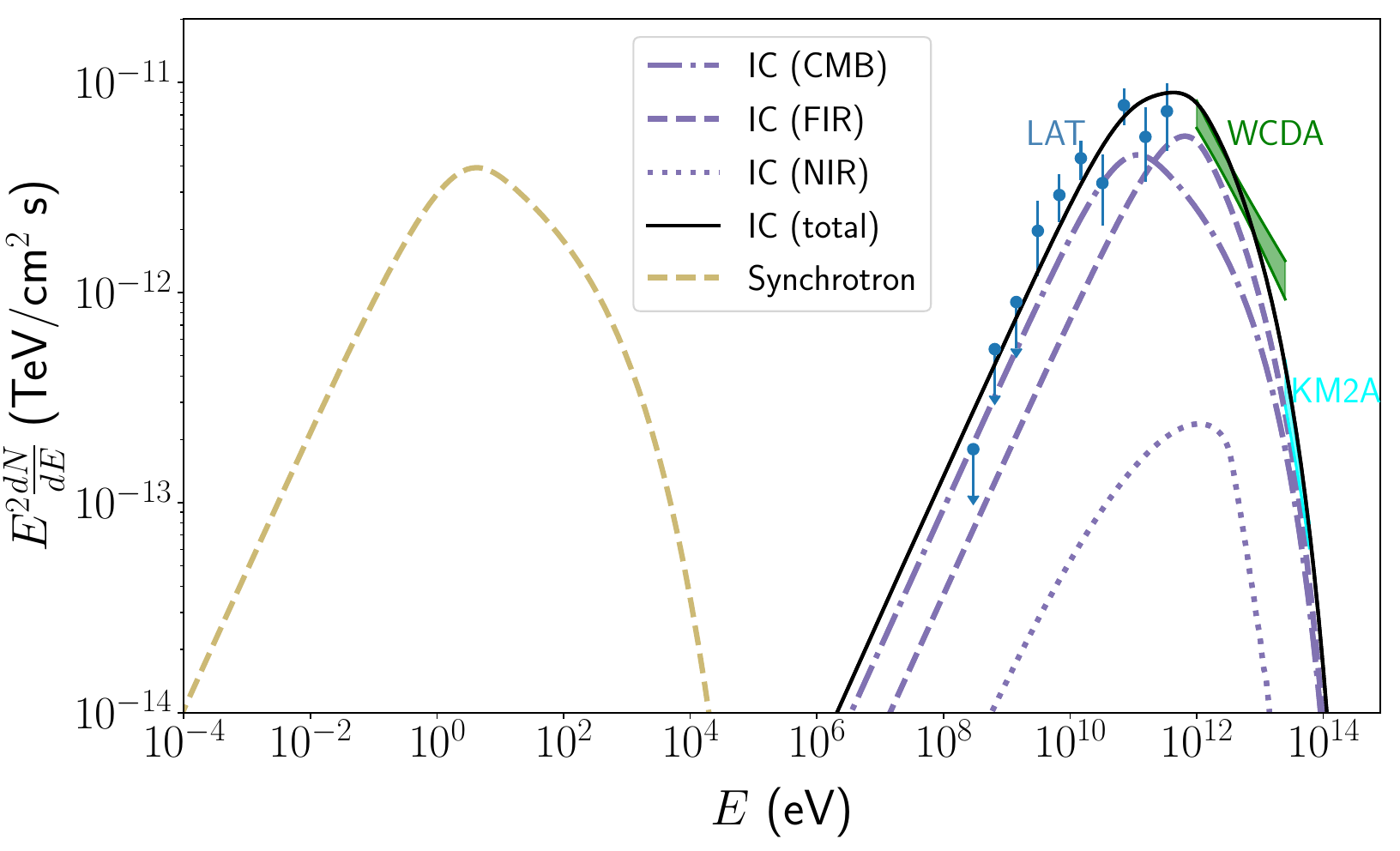}
    \includegraphics[width=0.7\textwidth]{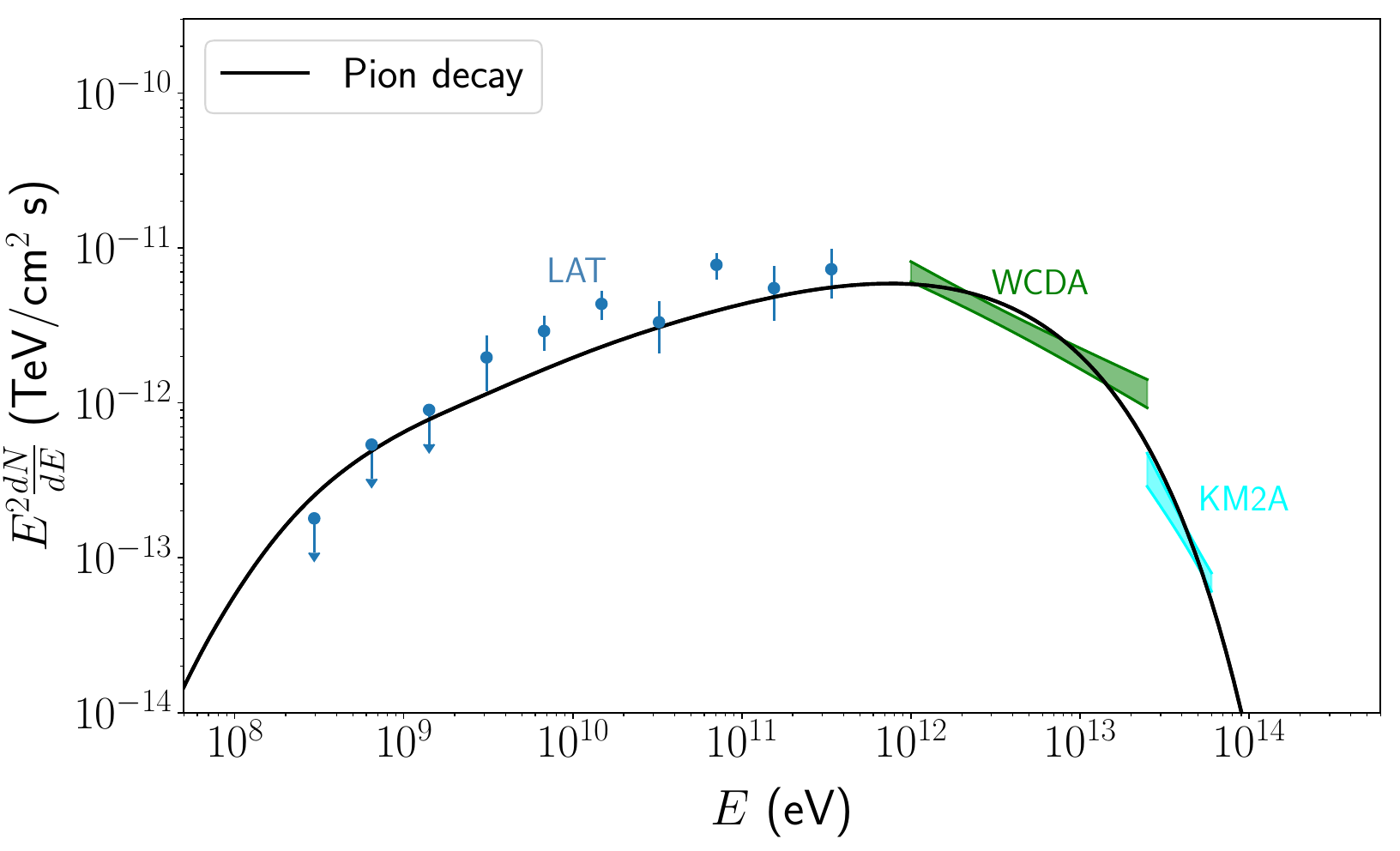}
    \caption{\textit{Fermi}-LAT data points from this work with WCDA and KM2A uncertainty bands. \textit{Top}: Fit from the leptonic scenario including the contributions to the IC emission from the three photon fields considered, and the predicted synchrotron emission for a magnetic field of 3~$\mu$G. \textit{Bottom}: hadronic scenario.}
    \label{sed}
\end{figure}

We found that an electron energy distribution following a broken power-law with an exponential cutoff can describe the data reasonably well. The spectral indices below and above the break energy are $1.67_{-0.14}^{+0.15}$ and $3.34_{-0.19}^{+0.06}$, the break energy is $3.7_{-0.7}^{+0.4}$~TeV and the particle cutoff energy is $51_{-10}^{+5}$~TeV ($1\sigma$ statistical uncertainties are given). The parameters obtained using the alternative local interstellar radiation fields from \cite{2018A&A...617A..78T} are $1.94_{-0.22}^{+0.29}$, $3.09_{-0.29}^{+0.21}$, $4.8_{-1.2}^{+1.8}$~TeV and  $36_{-7}^{+9}$~TeV, respectively. Both sets are compatible within the uncertainties. The total energy in the leptons is $3.5\times10^{46}$~erg for a source distance of 0.65~kpc. The data points and the leptonic model are shown in figure \ref{sed}, where we also plot the predicted synchrotron emission from the same relativistic electrons for a magnetic field of 3~$\mu$G, the average field value in the interstellar medium. The synchrotron peak is predicted to show at a photon energy of $1-10$~eV for magnetic field values of $1-6\,\mu$G.

For a hadronic scenario the fit to the data is also good. We chose a particle energy distribution described by a power-law with an exponential cutoff. The particle spectral index required is $1.64_{-0.08}^{+0.09}$ and the cutoff energy is $38_{-3}^{+5}$~TeV, for a total energy in the particles of $8.4\times 10^{48}$~erg~$\times \left(\frac{1}{n} \right)$, with $n$ the number density of the target gas in units of cm$^{-3}$ (again for a source distance of 0.65~kpc).

The total energy in the particles is reasonable for both scenarios since it represents a small fraction of the typical kinetic energy in a supernova explosion ($\sim 10^{51}$~erg). The particle distribution in the leptonic scenario is not unlike those of other supernova remnants. Broken power-laws are observed in the radio spectra of SNRs where the break could be associated to the maximum energy of accelerated electrons or to synchrotron cooling. If the SNR is relatively evolved with an age of $\sim 10$~kyr and this age is of the order of the synchrotron cooling time of 4~TeV-electrons (the break energy in our leptonic scenario), the required magnetic field is $\sim 18\,\mu$G \citep{1970RvMP...42..237B}, which is reasonable for an SNR. However, given the lack of a synchrotron counterpart for the source in low energy catalogs, the magnetic field is likely lower, perhaps of the order of $3\,\mu$G. In that case the particle break would imply an unreasonable age of $\sim 300$~kyr for the SNR. This could mean that the observed break is not related to electron cooling and considering other sources of cooling such as from IC emission might also be insufficient. More complex models taking into account other scenarios and the evolution of an SNR should be explored in the future to explain the particle spectrum.

Regarding the hadronic scenario the required particle energy distribution below the cutoff is of the form $E^{-1.64}$, much harder than typically observed in hadronic emission from SNRs \citep[e.g.,][]{2013Sci...339..807A,2016ApJ...816..100J} and harder than predictions from standard shock acceleration theory \citep{1978ApJ...221L..29B}. However, such a hard particle spectrum could result from cosmic ray illumination of nearby clouds \citep[e.g.,][]{2007ApJ...665L.131G} if they existed in the region. We leave for the future a deeper study of the interstellar gas at the location of \source as well as more detailed modeling, for example involving different particle populations in the SNR or a mixed leptonic-hadronic model. More observations will also be necessary to confirm or discard the SNR scenario for \source.

\subsection{Pulsar wind nebula hypothesis}
A PWN is formed by winds powered by an energetic pulsar. In the $\gamma-$ray energy range the radiation from the PWN comes from relativistic electron and positron pairs in the form of IC emission. In this section we explore the possibility that the GeV$-$TeV emission results from a pulsar, either one of the pulsars known to exist in the direction of \source or a yet undiscovered pulsar, and we derive the physical parameters required.

\cite{2018A&A...612A...2H} carried out a detailed study of the TeV population of PWN and showed that most of the $1-10$~TeV luminosities range from $\sim 10^{33}$ to $\sim 2\times 10^{35}$~erg~s$^{-1}$. We calculated the $1-10$~TeV energy flux from \source using the WCDA spectrum. For a luminosity of $10^{33}$~erg~s$^{-1}$ the measured flux implies a source distance of $\sim0.8$~kpc. On the other hand, the radius of the PWN is likely not larger than the intrinsic radius of the largest known PWN, HESS~J$1825-137$, of $\sim75$~pc \citep[e.g.,][]{2020A&A...640A..76P}, and therefore the largest possible distance to \source in the PWN scenario is likely $\sim3$~kpc. We adopted a reference distance of 1~kpc in this model.

We took some features of a commonly used simple model \citep[described for example by][]{2018A&A...612A...2H}. The injection of energy into the PWN is determined by the energy loss rate of the pulsar, $\dot{E}$, which evolves in time as \citep[e.g.,][]{1973ApJ...186..249P} $$\dot{E}(t) = \dot{E}_0\left(1+\frac{t}{\tau_0}\right)^{-\frac{n+1}{n-1}}$$ with $n=3$ according to a pure dipole magnetic field. Here, $\tau_0$ is the initial spin-down timescale and $\dot{E}_0$ the initial spin-down luminosity of the pulsar. We also allowed the values of $\dot{E}_0$ and $\tau_0$ to vary, as well as the age of the pulsar.

Relativistic electrons and positrons are injected into the PWN during a time equal to the age of the pulsar subsequently cooling off by emitting synchrotron and IC photons. We calculated the particle differential energy spectrum, $N(E,t)$, from the energy-loss equation \citep[see, e.g.,][]{2014JHEAp...1...31T}, $$\frac{\partial N}{\partial t} = Q(E,t) - \frac{\partial}{\partial E}(bN)$$ where $b=b(E,t)$ is the energy loss rate of the particles, which we assumed to be dominated by radiative losses, and $Q(E,t)$ is the spectrum of particles injected into the system per unit time, which we assumed to be a broken power-law with an exponential cutoff, having spectral indices $s_1$ and $s_2$ before and after the break energy, $E_{br}$: $$ Q(E,t)=Q_0(t)\left\lbrace\begin{array}{c} (E/E_{br})^{-s_1}\,\,\,\,\mbox{if $E \leq E_{br}$} \\ (E/E_{br})^{-s_2} \,\,\,\,\mbox{if $E>E_{br}$} \end{array}\right.$$ This is justified by observations of many PWN \citep[see, e.g.,][]{2006ARA&A..44...17G}. The injected spectrum is normalized so that the energy content in the particles is a fraction $\eta$ of the spin-down luminosity, $$\eta \dot{E}(t) = \int_{E_{min}}^{E_{max}} E\,Q(E,t)dE,$$ with $\eta$ a free parameter, as well as $s_1$, $s_2$ and $E_{br}$. We set the values of $E_{min}$ and $E_{max}$  to 100~MeV and 500~TeV, respectively.

We used the {\tt GAMERA} package \citep{Hahn:2016CO} to solve the energy-loss equation, and injected the particle populations as a series of bursts. Each population evolves until the desired age of the system is reached, at which point the synchrotron and $\gamma-$ray fluxes from the particles are calculated. As in the previous section, the IC emission is calculated with the full Klein-Nishina cross-section \citep[e.g.,][]{1970RvMP...42..237B} and the same seed photon fields.

The magnetic field in which the particles evolve is given as a function of time by $$B(t) = \frac{B_0}{1+\left(\frac{t}{\tau_0}\right)^\alpha}$$ with $\alpha=0.5$ following \cite{2008ApJ...676.1210Z}, and the initial magnetic field $B_0$ is also a free parameter of the model. We put a further constrain regarding the value of $B_0$ such that the total energy content in the magnetic field within the volume of the TeV emission region (modeled as a sphere) at a time equal to the age of the system is a fraction $1-\eta$ of the total spin-down energy injected by the pulsar.

From the WCDA flux, $F$, measured from \source in the energy range $1-10$~TeV, and the source angular radius ($\theta=1.29\degr$) we calculated the surface brightness which is independent of the distance to the source, $$S = \frac{L}{4\pi R^2} = \frac{F}{\tan^2\theta} \approx 2.8\times10^{29}\,\mbox{erg~pc$^{-2}$~s$^{-1}$},$$ where $L$ and $R$ are the luminosity in the same energy range and the source radius. \cite{2018A&A...612A...2H} derived a set of values for the parameters of their model such as the initial spin-down power of a pulsar, the initial spin-down timescale, the initial magnetic field in the nebula, and others, to reproduce the average trends obtained from observations (the baseline model). They also changed these parameters in a range of values to explain the scatter of measured PWN observables (varied models). For our estimated surface brightness value, the baseline model from \cite{2018A&A...612A...2H} predicts the spin-down luminosity of the associated pulsar at the current epoch to be $\sim 7\times 10^{34}$~erg~s$^{-1}$, although in their varied models this value could be as large as $\sim 5\times 10^{36}$~erg~s$^{-1}$. We thus kept the current pulsar spin-down luminosity below $5\times 10^{36}$~erg~s$^{-1}$ in the fits.

We also set the (apparent) TeV efficiency, $\epsilon_{\mbox{\tiny TeV}}$, defined as the ratio of the $1-10$~TeV luminosity and the current spin-down luminosity of the pulsar to be lower than 0.1, following the trends seen in \cite{2018A&A...612A...2H}. This has implications regarding the age for a given value of $\tau_0$ since for an older system the required spin-down luminosity is lower meaning that $\epsilon_{\mbox{\tiny TeV}}$ increases.

\begin{figure}
    \centering
    \includegraphics[width=0.7\textwidth]{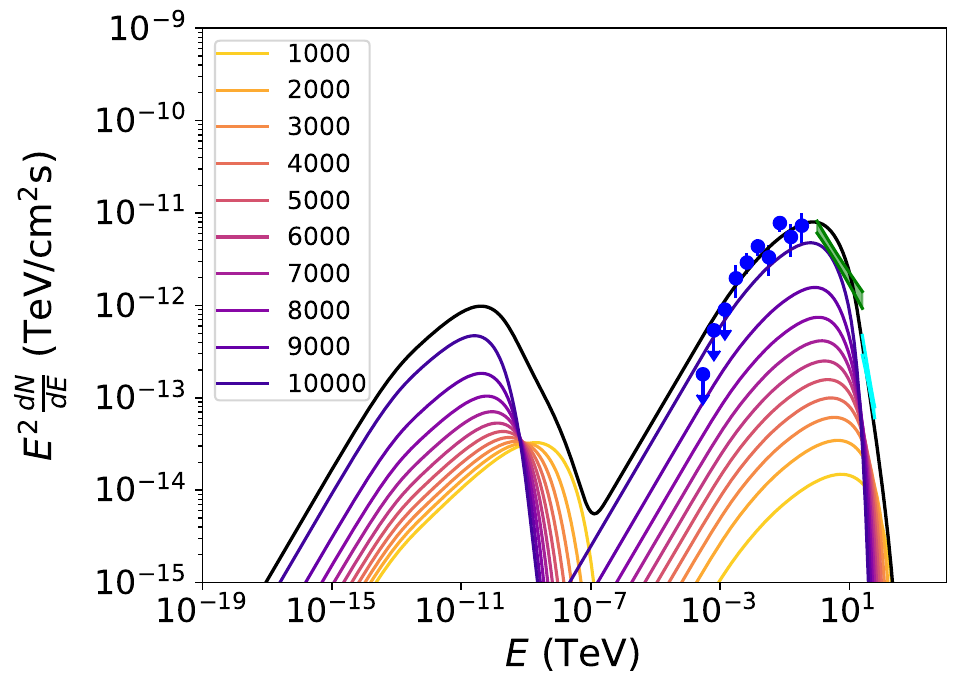}
    \caption{Data from Fig. \ref{sed} and the PWN model explained in the text (solid lines). The colours represent the contributions from each of the particle populations that are injected throughout the history of the system (whose age is labeled in yr) while the black line is their sum.}
    \label{fig:PWNmodel}
\end{figure}

Our derived set of parameter values can be seen in table \ref{table2} and the SEDs in figure \ref{fig:PWNmodel}. As can be seen in the figure, the highest energy photons measured by KM2A are produced by the youngest lepton populations while the LAT fluxes are mostly accounted for by the accumulation of old leptons in the system. We note that all the catalogued pulsars in the region, some of which were mentioned in section \ref{sec:intro} and are shown in figure \ref{fig:tsmap}, are located at large distances ($> 4$~kpc) according to their dispersion measure. The closest known pulsar at $\sim4.3$~kpc is PSR~J$1946+24$, a distance which would make the TeV source unreasonably large. Also, as noted earlier, this pulsar has a very long period and thus likely a very low spin-down luminosity. In our PWN model for \source a yet unknown pulsar must be responsible for the emission. Our simple model predicts some of its properties and it could be tested with future observations.

\begin{table*}
\centering
\caption{Parameters of the PWN model for \source (the source distance is fixed to 1~kpc)}
\begin{tabular}{|l|c|}
\hline
\hline
Parameter & \\
\hline
Pulsar age ($t$, kyr) & 10 \\
Initial spin-down luminosity ($\dot{E}_0$, erg~s$^{-1}$) & $3\times 10^{36}$\\
Current spin-down luminosity ($\dot{E}$, erg~s$^{-1}$) & $2.5\times 10^{34}$ \\
Initial spin-down timescale ($\tau_0$, kyr) & 1\\
Particle spectral index before the break ($s_1$) & 1.8 \\
Particle spectral index after the break ($s_2$) & 2.3 \\
Particle break energy ($E_{br}$, TeV) & 0.7 \\
Particle cutoff energy (TeV) & 200 \\
Initial magnetic field value ($B_0$, $\mu$G) & 18\\
Current magnetic field value ($\mu$G) & 4.3\\
Fraction of the spin-down luminosity going to leptons ($\eta$) & 0.2\\
\hline
\end{tabular}\\
\label{table2}
\end{table*}

Finally, we note that our model does not attempt to explain the source morphology or extension. In a PWN scenario for the $\gamma-$rays the GeV emission is expected to be more extended than the TeV emission since the GeV-emitting particles can propagate farther away before losing their energy. In the case of \source, although the KM2A source is more compact than the WCDA source, the latter is more extended than the GeV component. This could be due to several factors. For example low energy particles could have escaped the system and their emission is not detected due to the sensitivity limit of the LAT. Limits in instrument sensitivities, poor statistics, low angular resolution and a possible intrinsic complex source morphology could partly explain the inconsistencies in the positions obtained from different detectors. It is also possible that two or more unrelated sources are being detected by the LAT and LHAASO in the same line of sight. Deeper observations in different wavelengths will be needed to understand the origin of the $\gamma-$rays.

\section{Summary}
We discovered an extended GeV source at the location of the TeV source \source observed by LHAASO, using LAT data. The GeV emission shows a hard photon spectrum that connects smoothly with that above 1~TeV. We explored an SNR model with simple hadronic and leptonic one-population scenarios. The hadronic scenario requires a particle distribution with a harder spectral index than predicted by linear shock acceleration theory, while in the leptonic scenario more work is needed to understand the features of the particle spectrum such as the energy break. A simple PWN model for the origin of the $\gamma-$rays gives reasonable results and predicts the existence of a pulsar with a spin-down luminosity of $\sim 10^{34}$~erg~s$^{-1}$ responsible for the emission. More detailed observations and modeling are necessary to reveal the nature of this very-high-energy source.

\section*{Acknowledgements}
We thank the anonymous referee for comments that helped improve this work. We thank funding from Universidad de Costa Rica under grant B8267. This research has made use of NASA's Astrophysics Data System.

\section*{Data availability.} This work makes use of publicly available \textit{Fermi}-LAT data provided online by the Fermi Science Support Center at \\ http://fermi.gsfc.nasa.gov/ssc/, as well as publicly available data from the Large High Altitude Air Shower Observatory first source catalog.

\bibliographystyle{mnras}
\bibliography{lhaasoSrc}

\bsp    
\label{lastpage}

\end{document}